\documentclass[11pt]{article}
\def\be{\begin{equation}}
\def\ee{\end{equation}}
\def\ba{\begin{eqnarray}}
\def\ea{\end{eqnarray}}
\def\nn{\nonumber}
\def\lb{\label}
\def\dfrac{\displaystyle\frac}
\def\bb{\bibitem}
\def\E{{\cal E}}

\begin{document}
\begin{titlepage}
\date{}
\title{\begin{flushright}\begin{small}    LAPTH-024/17
\end{small} \end{flushright} \vspace{1.5cm}
On the Smarr formula\\ for rotating dyonic black holes}
\author{G\'erard Cl\'ement$^a$\thanks{Email: gclement@lapth.cnrs.fr},
Dmitri Gal'tsov$^{b,c}$\thanks{Email: galtsov@phys.msu.ru} \\ \\
$^a$ {\small LAPTh, Universit\'e Savoie Mont Blanc, CNRS, 9 chemin de Bellevue,} \\
{\small BP 110, F-74941 Annecy-le-Vieux cedex, France} \\
$^b$ {\small Department of Theoretical Physics, Faculty of Physics,}\\
{\small Moscow State University, 119899, Moscow, Russia }\\
$^c$ {\small  Kazan Federal University, 420008 Kazan, Russia}}

\maketitle

\begin{abstract}
We revisit the derivation by Tomimatsu of the generalized Komar
integrals giving the mass and angular momentum of rotating
Einstein-Maxwell black holes. We show that, contrary to Tomimatsu's
claim, the usual Smarr formula relating the horizon mass and angular
momentum still holds in the presence of both electric and magnetic
charges. The simplest case is that of dyonic Kerr-Newman black
holes, for which we recover the modified Smarr formula relating the
asymptotic mass and angular momentum, the difference between
asymptotic and horizon masses being equal to the sum of the two
Dirac string masses. Our results apply in particular to the case of
dyonic dihole solutions which have been investigated recently.
\end{abstract}
\end{titlepage}
\setcounter{page}{2}

\setcounter{equation}{0}
\section{Introduction}
The Smarr formula \cite{smarr} relating the mass, angular momentum,
entropy and electric charge of black holes was originally designed
for the electrically charged Kerr-Newman solution. Its possible
generalizations were extensively analyzed by Carter \cite{carter} on
the basis of Komar integrals \cite{komar}. More general solutions
describing axisymmetric configurations of multiple rotating black
holes (possibly joined by strings) endowed with electric and also
magnetic charges were discussed recently. These solutions usually
have a simple description in terms of Ernst potentials, while the
metric and the electromagnetic potentials are rather complicated.
For such situations Tomimatsu designed his original formulas
\cite{tom83,tom84} expressing black hole parameters in terms of both
the metric variables and the Ernst potentials taken on the symmetry
axis. These formulas suggested in 1984 were successfully applied for
multiple electrically charged rotating black holes \cite{manko2013}.

On the other hand, when the Tomimatsu formulas (completed by an
analogous expression for magnetic charge) were applied to
multi-dyons
\cite{tom84,cabrera2013,cabrera2014,cabrera2015,manko2015}, it was
observed that the resulting values for the black hole parameters
failed to obey the standard Smarr relation, but obeyed a generalized
Smarr relation with both electric and magnetic contributions.
However, the derivation by Tomimatsu \cite{tom84} gives little
details of the underlying calculations, so to clarify the situation
a new derivation is necessary. Here such a derivation is presented,
showing that in the original Tomimatsu formulas an important term is
missing. Correcting the Tomimatsu formulas, we obtain a new version
which, when applied to dyons, leads to the standard Smarr relation
between the local horizon mass, angular momentum and electric
charge.

In passing we establish the crucial role played by the Dirac strings
associated with magnetic monopoles in the mass and angular momentum
balance equations. We show that for the Kerr-Newman solution with
both electric and magnetic charges the Dirac strings are endowed
with non-zero generalized Komar masses which should be taken into
account in the Smarr formula for the total mass. We also find that
the symmetric choice of gauge for the vector potential (with both
North and South pole Dirac strings present with equal weights) for
dyons is essential to achieve the total angular momentum balance of
the configuration.

\setcounter{equation}{0}
\section{Generalized Komar mass and angular momentum}
We first review the generalized Komar formulas \cite{carter} giving
the masses and angular momenta of extended sources of
Einstein-Maxwell fields. The Komar mass and angular momentum for an
asymptotically flat, stationary, axisymmetric configuration are
given by the integrals over a spacelike 2-surface at infinity
\cite{komar}\footnote{We use the metric signature (-+++) and the
convention $d\Sigma_{\mu\nu}=1/2
\sqrt{|g|}\,\epsilon_{\mu\nu\lambda\tau} dx^\lambda dx^\tau$ with
$\epsilon_{t\rho z\varphi}=1$ in Weyl coordinates. We will label $t,
\varphi$ by an index $a$, and the remaining coordinates $\rho, z$ by
$i,j$. In Sect. 4 we will also use prolate spheroidal coordinates
$x,y$ instead of $\rho, z$. The two-dimensional Levi-Civita symbol
$\epsilon_{ij}$ is defined with $\epsilon_{\rho z}=1$ and
$\epsilon_{xy}=1$ respectively.}:
 \ba
M &=& \frac1{4\pi}\oint_\infty D^\nu k^{\mu}d\Sigma_{\mu\nu}, \lb{koM}\\
J &=& -\frac1{8\pi}\oint_\infty D^\nu m^{\mu}d\Sigma_{\mu\nu}
\lb{koJ}
 \ea
where $k^\mu = \delta^\mu_t$ and $m^\mu = \delta^\mu_\varphi$ are
the Killing vectors associated with time translations and rotations
around the $z$-axis.

Because the integrand $D^\nu k^{\mu}$ is antisymmetric, one can
apply the Ostrogradsky theorem to transform
 \be
M = \sum_n\dfrac1{4\pi}\oint_{\Sigma_n} D^\nu k^\mu d\Sigma_{\mu\nu}
+ \frac1{4\pi}\int D_\nu D^\nu k^\mu dS_\mu, \lb{koM1}
 \ee
where $\Sigma_n$ are the spacelike surfaces bounding the various
sources, and the second integral is over the bulk. Using again the
fact that $k$ is a Killing vector and the Einstein equations, we obtain
 \be
D_\nu D^\nu k^\mu = -[D_\nu,D^\mu]k^\nu = - {R^\mu}_\nu k^\nu =
-8\pi{T^\mu}_\nu k^\nu.
 \ee
Here ${T^\mu}_\nu$ is the electromagnetic energy-momentum tensor
 \be
{T^\mu}_\nu = \dfrac1{4\pi}\left[F^{\mu\rho}F_{\nu\rho} -
\dfrac14\delta^\mu_\nu F^{\rho\sigma}F_{\rho\sigma}\right],
 \ee
with $F_{\mu\nu} = \partial_\mu A_\nu - \partial_\nu A_\mu$. It
follows that the bulk contribution to the Komar mass (\ref{koM1}),
which after Tomimatsu \cite{tom84} we will call $M^E$, may be
transformed to
 \ba\lb{ME}
M^E &\equiv& \frac1{4\pi}\int D_\nu D^\nu k^\mu dS_\mu = -2\int {T^t}_t \sqrt{|g|}d^3x \nn\\
&=& -
\frac1{4\pi}\int\left(F_{it}F^{it}-F_{i\varphi}F^{i\varphi}\right)\sqrt{|g|}d^3x
\nn\\ &=& -\frac1{4\pi}\int\partial_i\left[\sqrt{|g|}\left(A_tF^{it}
-A_\varphi F^{i\varphi}\right)\right]d^3x \nn\\
&=& \sum_n \frac1{4\pi}\oint_{\Sigma_n}\left(A_t F^{it}-A_\varphi
F^{i\varphi}\right)d\Sigma_i,
 \ea
where we have used the Maxwell equations in the bulk outside the
sources, and again the Ostrogradsky theorem. Note that we have
implicitly assumed in the last step that
 \be\lb{potinf}
\oint_{\infty}\left(A_t F^{it}-A_\varphi
F^{i\varphi}\right)d\Sigma_i = 0.
 \ee
Were this condition not satisfied, the final result (\ref{ME})
should also include a surface integral at infinity.

Returning to (\ref{koM}), we can write the total mass as the sum of
the masses of the individual sources
 \be\lb{Mtot}
M = \sum_n M_n,
 \ee
with
 \be\lb{Mn}
M_n = \frac1{8\pi}\oint_{\Sigma_n}\left[g^{ij}g^{ta}\partial_jg_{ta}
+2(A_t F^{it}-A_\varphi F^{i\varphi})\right]d\Sigma_i,
 \ee
where the first term may be viewed as the gravitational contribution
to the source mass, and the second term as the electromagnetic
contribution.

Similarly, (\ref{koJ}) can be transformed to
 \be
J  = -\sum_n\dfrac1{8\pi}\oint_{\Sigma_n} D^\nu m^\mu
d\Sigma_{\mu\nu} + \int {T^\mu}_\nu m^\nu dS_\mu, \lb{koJ1}
 \ee
and the second, bulk contribution $J^E$ can be further transformed
to
 \ba
J^E &=& \frac1{4\pi}\int F_{i\varphi}F^{it}\sqrt{|g|} \,d^3x\nn\\
&=&  \frac1{4\pi}\int\partial_i\left(\sqrt{|g|} A_\varphi F^{it}
\right)d^3x \nn\\ &=& -\sum_n \frac1{4\pi}\oint_{\Sigma_n}A_\varphi
F^{it}\,d\Sigma_i,
 \ea
under the assumption
 \be\lb{potinf1}
\oint_{\infty}A_\varphi F^{it}\,d\Sigma_i = 0.
 \ee
The total angular momentum then decomposes as
 \be\lb{Jtot}
J = \sum_n J_n, \qquad J_n =
-\frac1{16\pi}\oint_{\Sigma_n}\left[g^{ij}g^{ta}
\partial_jg_{\varphi a} +4A_\varphi F^{it}\right]d\Sigma_i.
 \ee

These formulas give the masses and angular momenta as fluxes through
surfaces, and thus necessitate the knowledge of the gravitational
and electromagnetic potentials off these surfaces. In the case where
$\Sigma_n$ is the horizon of a rotating black hole, Tomimatsu
\cite{tom84} derived formulas which have the advantage of involving
only potentials on-shell, that is on the horizon.

\setcounter{equation}{0}
\section{Correcting the Tomimatsu formulas}
Now we revisit the derivation of the mixed formulas for the mass and
angular momentum of rotating black holes given by Tomimatsu in
\cite{tom84}. We call these ``mixed'' because they involve both
physical metric and Maxwell field components, and Ernst potentials.
We will spell out a number of details which were omitted in the
rather elliptic derivation of \cite{tom84}, with the conclusion that
the Tomimatsu formula for the black hole mass should be corrected.
The use of an incorrect formula has led Tomimatsu himself, and other
authors \cite{cabrera2013}-\cite{manko2015} to the incorrect
conclusion that the usual Smarr mass formula should be modified when
magnetic charges are present.

Before proceeding we need to recall the definition of the complex
Ernst potentials, which will be used in the following. In Weyl
coordinates, these are defined by
 \be
{\cal E} = F - \overline{\psi}\psi + i \chi\,, \qquad \psi = v +
iu\,,
 \ee
where the electric and magnetic scalar potentials $v$ and $u$ are
such that
 \be\lb{stat2}
v = A_t, \qquad \partial_i u = \rho\,\epsilon_{ij}{F_j}^\varphi,
\end{equation}
with $x^1=\rho$, $x^2=z$, and the twist potential $\chi$ is defined
by
 \be\lb{twist}
\partial_i\chi = -F^2\rho^{-1}\epsilon_{ij}\partial_j\omega
+ 2(u\partial_i v - v\partial_i u)\,.
\end{equation}

\subsection{Mass}
First consider the generalized Komar mass formula (\ref{Mn}), where
$\Sigma_n$ is the horizon $H$ of a rotating black hole, and the
spacetime metric is written in the standard Weyl form
 \be\lb{weyl}
ds^2 = -F(dt-\omega d\varphi)^2 +
F^{-1}[e^{2k}(d\rho^2+dz^2)+\rho^2d\varphi^2].
 \ee
The horizon corresponds to $N^2 \equiv \rho^2/g_{\varphi\varphi} =
0$ with
 \be
g_{\varphi\varphi} = F^{-1}\rho^2-F\omega^2 > 0,
 \ee
so that $H$ is a cylindrical surface $\rho=0$, $t=$ constant, on which
$\sqrt{|g|}g^{\rho\rho}=\rho=0$, and $\omega$ takes a constant value
$\omega_H = \Omega_H^{-1}$, with $\Omega_H$ the horizon angular
velocity. The horizon mass (\ref{Mn}) decomposes as $M_H = M_H^G +
M_H^E$, with the gravitational contribution
 \ba\lb{MG1}
M_H^G &=&
\frac1{8\pi}\int_{H}\sqrt{|g|}g^{\rho\rho}g^{ta}\partial_jg_{ta}\,dzd\varphi
\nn\\ &=& \frac1{8\pi}\int_{H}\left[\rho F^{-1}\partial_\rho
F + \rho^{-1}F^2\omega\partial_\rho\omega\right]dzd\varphi \nn\\
&=& \frac1{8\pi}\int_{H}\omega\left[\partial_z\chi + 2(v\partial_z u
- u\partial_z v)\right]dzd\varphi,
 \ea
where we have discarded the first term of the integrand in the
second line, which (assuming $F_H\neq0$) vanishes on the horizon
$\rho=0$, and used the definition (\ref{twist}) of the Ernst twist
potential $\chi = {\rm Im}\E$ to express the second term in terms of
Ernst potentials. The electromagnetic contribution is
 \ba\lb{ME1}
M_H^E &=& \frac1{4\pi}\int_H\sqrt{|g|}\left(A_tF^{\rho t} -
A_\varphi F^{\rho\varphi}\right)dzd\varphi \nn\\
&=& \frac1{4\pi}\int_H\sqrt{|g|}\left(\omega v -
A_\varphi\right)F^{\rho\varphi}\,dzd\varphi,
 \ea
on account of $\sqrt{|g|}(F^{\rho t} - \omega F^{\rho\varphi}) =
\sqrt{|g|}g^{\rho\rho}\omega^{-1}F^{-1} F_{\rho\varphi} = 0$ on the horizon.
Using the electromagnetic duality equation (\ref{stat2}) this can be
transformed to
 \ba\lb{ME2}
M_H^E &=& -\frac1{4\pi}\int_H\left(\omega v -
A_\varphi\right)\partial_z u\,dzd\varphi \nn\\
&=& -\frac1{4\pi}\int_H\left[\omega\left(v\partial_z u - u
\partial_z v\right) - \partial_z(uA_\varphi)\right]dzd\varphi
 \ea
where we have used the constancy over the horizon of
 \be
A_\varphi+\omega v = -\omega_H\Phi_H,
 \ee
with
 \be\lb{phiH}
-\Phi_H= A_t+\Omega_H A_\varphi
 \ee
the horizon electric potential in the horizon co-rotating frame
\cite{carter}. Adding (\ref{MG1}) and (\ref{ME1}), we obtain
 \be\lb{MH}
M_H = \frac1{8\pi}\int_H\left[\omega\partial_z{\rm Im}\E +
2\partial_z(A_\varphi\,{\rm Im}\psi)\right]dzd\varphi,
 \ee
which differs from Tomimatsu's \cite{tom84} Eq. (52) by the presence
of the second term. We will discuss the consequences of this
difference shortly.

\subsection{Angular momentum}
Similarly, the first, gravitational contribution to the horizon
angular momentum is
 \ba
J^G_H &=& -\frac1{16\pi}\int_H\sqrt{|g|}g^{\rho\rho}g^{ta}
\partial_\rho g_{\varphi a}\,dzd\varphi \nn\\
&=& -\frac1{16\pi}\int_{H}\left[2\omega(1-\rho
F^{-1}\partial_\rho F) - \rho^{-1}(\rho^2+F^2\omega^2)\partial_\rho\omega\right]dzd\varphi \nn\\
&=&
-\frac1{16\pi}\int_{H}\left\{2\omega-\omega^2\left[\partial_z\chi +
2(v\partial_z u - u\partial_z v)\right]\right\}dzd\varphi.
 \ea
The electromagnetic contribution is
 \ba
J^E_H
&=& -\frac1{4\pi}\int_H\sqrt{|g|}A_\varphi F^{\rho t}\,dzd\varphi \nn\\
&=& \frac1{4\pi}\int_H\omega A_\varphi \partial_z u dzd\varphi \nn\\
&=& \frac1{8\pi}\int_H\omega\left[(A_\varphi+\omega v)\partial_z u -
\omega v\partial_z u + u(\omega\partial_z v + \partial_z A_\varphi)
+ A_\varphi \partial_z u\right]dzd\varphi \nn\\
&=& \frac1{8\pi}\int_H\omega\left[(A_\varphi+\omega v)\partial_z u -
\omega(v\partial_z u - u\partial_z v) + \partial_z(A_\varphi u)
\right]dzd\varphi,
 \ea
where we have used the constancy of $A_\varphi+\omega v$ over the
horizon in the third line. Adding these two contributions together,
we obtain
 \be\lb{JH}
  J_H = \frac1{16\pi}\int_H\omega\left[-2 + \omega\partial_z{\rm
Im}\E + 2\partial_z(A_\varphi\,{\rm Im}\psi) - 2\omega
\Phi_H\partial_z{\rm Im}\psi\right]dzd\varphi,
 \ee
in agreement with Eqs. (54)-(55) of \cite{tom84}.

\subsection{The horizon Smarr formula}
To be self-contained, we first recall briefly the derivation of the
Tomimatsu formula giving the horizon electric charge \cite{tom84}.
In Weyl coordinates, this is defined by the flux
 \be
Q_H = \frac1{4\pi}\int_H\sqrt{|g|}F^{t\rho}dzd\varphi.
 \ee
The electric field is related to the Ernst potentials by
 \be\lb{elernst}
F^{ti} = \frac{g^{ij}}{g_{tt}}F_{tj} -
\frac{g_{t\varphi}}{g_{tt}}F^{\varphi i} = e^{-2k}\left[\partial_i v
+ \frac{F\omega}\rho\epsilon_{ij}\partial_ju\right],
 \ee
leading on the horizon $\rho=0$ to
 \be\lb{QH}
Q_H=\frac1{4\pi}\int_H \omega\partial_z{\rm Im}\psi\,dzd\varphi.
 \ee

Comparing (\ref{MH}) and (\ref{JH}), using the definitions of the
Hawking temperature $T_H=\kappa/2\pi$ and entropy $S={\cal A}_H/4$,
with the surface gravity $\kappa=e^{-k}/|\omega|$ and the horizon
area ${\cal A}_H=\int_H e^k|\omega|\,dzd\varphi$ (so that $T_HS =
\kappa{\cal A}_H/8\pi = \sigma/2$), and the value (\ref{QH}) of the
horizon electric charge, we recover the usual Smarr mass formula
 \be\lb{smarr}
M_H = 2\Omega_HJ_H + 2T_HS + \Phi_HQ_H.
 \ee
We have thus shown that this formula, which involves only electric
charges, is valid on the horizons for any asymptotically flat
configuration, including horizons carrying also magnetic charges.

As mentioned above, Tomimatsu obtained in \cite{tom84} an expression
$M'_H$ for the horizon mass which does not contain the second term
in (\ref{MH}). Consequently he obtained for his $M'_h$ a modified
Smarr formula (Eq. (57) of \cite{tom84}) which contains an
additional term ${M_A}^S$, equal to the opposite of the ``lost''
second term of (\ref{MH}), so as to account for the difference
$M'_H-M_H$. This extra term can contribute to the horizon mass in
the case of magnetically charged black holes, and the consequences
of this were explored in \cite{cabrera2013}-\cite{manko2015} in the
case of dyonic diholes, pairs of rotating black holes carrying
opposite electrical and magnetical monopole charges (so that the
global system has only dipole electromagnetic moments). However, as
we have shown, the usual Smarr relation holds in this case for the
horizon observables, provided they are correctly computed. It is
true that the two black holes are connected by a string with conical
singularity and carrying magnetic flux (Dirac string), but the
contributions of this string to the total mass and angular momentum
should be included as third separate contributions $M_{\rm string}$
and $J_{\rm string}$ (integrals on cylinders centered on the
strings) to the sums (\ref{Mtot}) and (\ref{Jtot}), without effect
on the horizon observables,

The situation is different in the case of the first and third
examples discussed in \cite{manko2015} (the dyonic Kerr-Newman
solution, and a system of two counter-rotating black holes with the
same electric and magnetic charges). In this case, not only will
some Dirac strings necessarily extend to infinity, but also the net
magnetic monopole potential
 \be
A_\varphi \sim -P(\cos\theta + C)
 \ee
(where $P$ is the net magnetic charge, and $C$ a constant governing
the strength of the Dirac strings) will go to constant values at
infinity for all $\theta$, so that the sums (\ref{Mtot}) and
(\ref{Jtot}) could include, besides the horizon and string
contributions, additional contributions from a surface at infinity.
We revisit the case of the dyonic Kerr-Newman black hole in the next
section.

\setcounter{equation}{0}
\section{The case of the dyonic Kerr-Newman black hole}
The metric and electromagnetic fields of this solution are given by
\cite{manko2015}:
 \ba\lb{KN}
F &=& \frac{f}\Sigma, \qquad  e^{2k} = \frac{f}{\sigma^2(x^2-y^2)},
\nn\\ f &=& \sigma^2(x^2-1) - a^2(1-y^2), \qquad \Sigma = (\sigma
x+M)^2 + a^2y^2, \nn\\ \omega &=& -a(1-y^2)\frac{2M(\sigma
x+M)-Q^2-P^2}f, \nn\\
A_t &=& \frac{-Q(\sigma x+M) + aPy}\Sigma, \quad A_\varphi = - Py -
C - aA_t(1-y^2),
 \ea
where the prolate spheroidal coordinates $x\ge1$, $y\in[-1,+1]$ are
related to the Weyl coordinates by
$\rho=\sigma(x^2-1)^{1/2}(1-y^2)^{1/2}$, $z=\sigma xy$, $\sigma$
being related to the mass $M$, electric charge $Q$, magnetic charge
$P$ and rotation parameter $a$ by $\sigma^2 = M^2-Q^2-P^2-a^2$. The
corresponding Ernst potentials are
 \be\lb{ernstKN}
\E = \frac{\sigma x - M + iay}{\sigma x + M + iay}, \qquad \psi =
\frac{-Q+iP}{\sigma x + M + iay}
 \ee
(our $\E$ is the complex conjugate of that of \cite{manko2015}, and
our $\psi$ is minus the complex conjugate of the $\Phi$ of
\cite{manko2015}), and their imaginary parts are
 \be\lb{chiuKN}
\chi={\rm Im}\E =\frac{2aMy}\Sigma, \quad u={\rm Im}\psi =
\frac{P(\sigma x+M) + aQy}\Sigma.
 \ee

First we observe that $A_tF^{rt}$ and $A_\varphi F^{r\varphi}$ (with
$r \sim \sigma x$) fall off at infnity more quickly than $1/r^2$, so
that the condition (\ref{potinf}) is satisfied. However the
situation is different for the angular momentum. Namely, in
transforming the Komar angular momentum (\ref{koJ}) into the sum
(\ref{Jtot}) we have dropped an integral over a large sphere at
spatial infinity parameterized by the angles $\theta$ (with
$y=\cos\theta$) and $\varphi$,
 \be
\frac1{4\pi}\int_\infty A_\varphi F^{rt}r^2 \sin\theta d\theta
d\varphi = \frac{PQ}2\int_{-1}^{+1}(y+C)dy = CPQ.
 \ee
So (\ref{Jtot}) is valid for dyons provided the constant $C$ ($-b_0$
in \cite{manko2015}) is set to zero. Let us emphasize that this
conclusion depends only on the asymptotic behaviour, and so
obviously extends to the case of axisymmetric dyonic multi-black
hole configurations, such as that discussed in Sect. IV of
\cite{manko2015}. For such configurations with net total electric
and magnetic charges $Q$ and $P$ both different from zero, horizon
and string angular momenta can consistently be defined only in the
gauge where the vector magnetic potential is asymptotically $A = -
P\cos\theta d\varphi$, so that that the asymptotic angular momentum
(\ref{koJ}) can be transformed into the sum (\ref{Jtot}).

The horizon is $x=1$. On the horizon we find from (\ref{KN})
$\omega_H=\Omega_H^{-1}$, with the horizon angular velocity
 \be
\Omega_H = \frac{a}{(M+\sigma)^2+a^2} = \frac{a}{\Sigma_0}.
 \ee
We use this to evaluate the integrals (\ref{QH}), (\ref{QH}),
(\ref{QH}). The evaluation of the horizon electric charge (\ref{QH})
leads naturally to
 \be
Q_H = \frac{\omega_H}2\int_{-1}^{+1}\partial_y u_H\,dy =
\frac{\omega_H}2\left[u(1,1)-u(1,-1)\right] = Q.
 \ee
The evaluation of the first term $1/8\pi\int_H\omega\partial_y u
dyd\varphi$ of the horizon mass results similarly, as in
\cite{manko2015}, in the value $M$. Concerning the second term, we
note that $A_\varphi=-Py$ on the horizon, so that only the part of
$u$ which is even in $y$ will contribute, leading to
 \be
\frac1{8\pi}\int_H2\partial_y(A_\varphi\,u)dyd\varphi = -\frac{P^2
(M+\sigma)}{\Sigma_0},
 \ee
and to the total horizon mass (defined \`a la Tomimatsu)
 \be\lb{MHKN}
M_H = M - \frac{P^2 (M+\sigma)}{\Sigma_0}\,.
 \ee
To compute the horizon angular momentun (\ref{JH}), we need to know
the horizon potential $\Phi_H$. From (\ref{phiH}) we obtain
\cite{manko2015}
 \be
\Phi_H = \frac{Q(M+\sigma)}{\Sigma_0}.
 \ee
We can then use (\ref{smarr}) to obtain
 \ba\lb{JHKN}
J_H &=& \frac{\omega_H}2\left[M_H-\sigma-\Phi_HQ_H\right] \nn\\ &=&
\frac{\omega_H}2\left[M-\sigma-\frac{(P^2+Q^2)(M+\sigma)}{\Sigma_0}\right]
= Ma = J.
 \ea

Comparing (\ref{MHKN}) and (\ref{JHKN}), we see that while the
(local) horizon mass and angular momentum satisfy the usual Smarr
formula (\ref{smarr}), the corresponding (global) quantities
evaluated at infinity satisfy a modified Smarr formula
\cite{manko2015} where the electric and magnetic charges stand on
equal footings:
 \be\lb{smarr1}
M = 2\Omega_HJ + 2T_HS + \Phi_HQ + \tilde{\Phi}_HP,
 \ee
with the horizon magnetic potential $\tilde{\Phi}_H =
P(M+\sigma)/\Sigma_0$.

A deeper issue is to understand the reason for the difference
between the horizon mass $M$ and the global mass $M_H$. This can
only be due to the contribution of the two Dirac strings ($y=1$ and
$y=-1$ with $x>1$) to the sum (\ref{Mtot}). Consider the integral
(\ref{Mn}) on a surface $\Sigma_{S_\pm}$ of equation
$y=\pm(1-\varepsilon)$, where the constant $\varepsilon$ shall be
taken to zero. Using
 \be
d\rho^2+dz^2 = \sigma^2(x^2-y^2)\left[\frac{dx^2}{x^2-1} +
\frac{dy^2}{1-y^2}\right],
 \ee
the flux (\ref{Mn}) through this surface is
 \be\lb{MS}
M_{S_\pm} = \mp
\frac1{8\pi}\int_{\Sigma_\pm}\sqrt{|g|}g^{yy}\left[g^{ta}\partial_yg_{ta}
+2(g^{ta}A_t\partial_yA_a - g^{\varphi
a}A_\varphi\partial_yA_a)\right]dxd\varphi
 \ee
($a=t,\varphi$), with $\sqrt{|g|}g^{yy}=1-y^2$. It follows that in
the limit $\varepsilon\to0$ ($y\to\pm1$), only the terms inside the
square brackets with a pole in $(1-y^2)$ will contribute to the
integral. The covariant metric tensor and electromagnetic vector
components remain finite in this limit, as well as the contravariant
metric components $g^{ab}$, with the exception of
$g^{\varphi\varphi} \sim F/\sigma^2(x^2-1)(1-y^2)$. So (\ref{MS})
reduces to
 \ba\lb{MS1}
M_{S_\pm} &=& \pm
\frac1{4\pi\sigma}\int_{y=\pm1}\frac{F}{x^2-1}A_\varphi\partial_yA_\varphi
dxd\varphi
\nn\\ &=& \pm\frac\sigma2\int_1^\infty\frac{Py(P-2aA_ty)}\Sigma dx \nn\\
&=& \frac{P}2\int_{\sigma+M}^\infty\frac{P(\xi^2-a^2)\pm
2aQ\xi}{(\xi^2+a^2)^2}\,d\xi
\nn\\
&=& \frac{P[P(M+\sigma)\mp aQ]}{2[(M+\sigma)^2+a^2]},
 \ea
where we have put $\xi=\sigma x+M$. The sum of the two string masses
 \be
M_{S_+} + M_{S_-} = \frac{P^2(M+\sigma)}{\Sigma_0},
 \ee
leads to
 \be
M_H + M_{S_+} + M_{S_-} = M,
 \ee
as required. A similar computation for the Dirac string angular
momenta yields consistently $J_{S_\pm} = 0$, as the contravariant
$g^{\varphi\varphi}$ does not occur in (\ref{Jtot}).

\setcounter{equation}{0}
\section{Conclusion}

Thus we solved the dilemma ``Tomimatsu vs Smarr'' in favor of the
latter and presented a corrected version of the Tomimatsu formulas
which reproduce the standard Smarr relation for the horizon mass of
dyons without an additional term associated with the magnetic
charge. Our modification should settle some problems which have been
raised in recent discussions of multi-dyonic black hole solutions of
the Einstein-Maxwell equations. We should mention that it has no
effect in the case of the recently presented quasiregular two-NUTty
dyonic black hole solution \cite{Clement:2017kas}, for which the
term omitted by Tomimatsu vanishes.

We have also discovered that Dirac strings give a non-zero
contribution to the total mass of the Kerr-Newman dyonic black hole,
so that Dirac strings are ``heavy''.  Taking them into account, one
finds that the Smarr relation for the total mass includes the
magnetic term. One unexpected feature concerns the gauge choice for
the vector potential of the magnetic charge. Usually it is believed
that by adding a constant term to $A_\varphi$ one can eliminate
either the North pole or the South pole Dirac strings without
affecting physics. This is not valid if gravity is taken into
account: only the symmetric gauge choice with both strings present
with equal weights leads to correct balance of mass and angular
momentum. This emphasizes the fact that adding a constant term to
$A_\varphi$ is a ``large'' gauge transformation affecting the
physical properties of the solution.

\section*{Acknowledgments} DG thanks LAPTh Annecy-le-Vieux for
hospitality at different stages of this work. DG also acknowledges
the support of the Russian Foundation of Fundamental Research under
the project 17-02-01299a and the Russian Government Program of
Competitive Growth of the Kazan Federal University.


\begin{thebibliography}{9}

\bb{smarr} L. Smarr, Phys. Rev. Lett. 30, 71 (1973).

\bb{carter} B. Carter, ``Black Hole Equilibrium States, Part 2:
General Theory of Stationary Black Hole States'', in Black Holes
(Les Houches 1972) ed B and C. DeWitt, (Gordon Breach, New York
1973)) 125-214.

\bb{komar} A. Komar, Phys. Rev. {\bf 113}, 934 (1959).

\bb{tom83} A. Tomimatsu, Progr. Theor. Phys. {\bf 70}, 385 (1983).

\bb{tom84} A. Tomimatsu, Progr. Theor. Phys. {\bf 72}, 73 (1984).

\bibitem{manko2013} V.~S.~Manko and R.~I.~Rabad\'an, Phys.Rev. D {\bf 89},
064049 (2014) [arXiv:1311.2326 [gr-qc]].

\bb {cabrera2013} I. Cabrera-Munguia, C. L\"ammerzahl, L.A. L\'opez
and A. Mac\'{\i}as, Phys. Rev. D {\bf 88}, 084062 (2013),
arXiv:1309.2556 [gr-qc].

\bb {cabrera2014} I. Cabrera-Munguia, C. L\"ammerzahl, L.A. L\'opez
and A. Mac\'{\i}as, Phys. Rev. D {\bf 90}, 024013 (2014)
[arXiv:1405.2629 [gr-qc]].

\bb {cabrera2015} I. Cabrera-Munguia, C. L\"ammerzahl and A.
Mac\'{\i}as,
Phys. Lett. B {\bf 743}, 357 (2015).

\bibitem{manko2015} V.~S.~Manko and H.~Garc\'{\i}a-Compe\'an, ``Remarks on
Smarr's mass formula in the presence of both electric and magnetic
charges'', arXiv:1506.03870 [gr-qc].

\bb{Clement:2017kas} G.~Cl\'ement and D.~Gal'tsov, Phys.\ Lett.\ B
{\bf 771}, 457 (2017), arXiv:1705.08017 [gr-qc].

\bibitem{Heusler:1997am}
  M.~Heusler,
  Phys.\ Rev.\ D {\bf 56}, 961 (1997)
  doi:10.1103/PhysRevD.56.961
  [gr-qc/9703015].

\bibitem{Cardenas:2016uzx}
  M.~Cardenas, O.~Fuentealba and J.~Matulich,
  JHEP {\bf 1605}, 001 (2016)
  doi:10.1007/JHEP05(2016)001
  [arXiv:1603.03760 [hep-th]].
\end{thebibliography}
\end{document}